\documentclass[aps,prl,epsfig,floats,twocolumn,superscriptaddress,amssymb,amsmath,showpacs]{revtex4}
\usepackage{graphicx}
\begin{document}
\title{Self-motile colloidal particles: from directed propulsion to random walk}

\author{Jonathan R. Howse}

\affiliation{Department of Physics and Astronomy, University of
Sheffield, Sheffield S3 7RH, UK}

\author{Richard A.L. Jones}
\email{r.a.l.jones@sheffield.ac.uk} \affiliation{Department of
Physics and Astronomy, University of Sheffield, Sheffield S3 7RH,
UK}

\author{Anthony J. Ryan}

\affiliation{Department of Chemistry, University of Sheffield,
Sheffield S3 7HF, UK}

\author{Tim Gough}

\affiliation{IRC in Polymer Engineering, University of Bradford, BD7
1DP, UK}

\author{Reza Vafabakhsh}

\affiliation{Institute for Advanced Studies in Basic Sciences,
Zanjan 45195-1159, Iran}

\author{Ramin Golestanian}
\email{r.golestanian@sheffield.ac.uk} \affiliation{Department of
Physics and Astronomy, University of Sheffield, Sheffield S3 7RH,
UK}

\date{\today}

\begin{abstract}
The motion of an artificial micro-scale swimmer that uses a chemical
reaction catalyzed on its own surface to achieve autonomous
propulsion is fully characterized experimentally. It is shown that
at short times, it has a substantial component of directed motion,
with a velocity that depends on the concentration of fuel molecules.
At longer times, the motion reverts to a random walk with a
substantially enhanced diffusion coefficient. Our results suggest
strategies for designing artificial chemotactic systems.
\end{abstract}

\maketitle

The directed propulsion of small scale objects in water is
problematic because of the combination of low Reynolds number and
Brownian motion on these length scales \cite{purcell1}.  In order to
achieve an artificial micro- or nano- scale swimmer that is able to
propel itself in a purposeful way, one needs both a swimming
strategy that works in the environment of low Reynolds number
\cite{taylor}, and a strategy for steering and directing the motion
that can overcome the ubiquity of Brownian motion. Common bacteria,
such as {\em E. Coli}, achieve propulsion by non-time-reversible
motion of long flagella, and employ a ``run and tumble'' strategy to
be able to swim towards or away from environmental stimuli
\cite{berg}.

One possibility for designing propulsion is to devise non-reciprocal
deformation strategies that are simple enough to be realizable
\cite{3SS}. Recently, an interesting example of such robotic
micro-swimmers has been made using magnetic colloids attached by
DNA-linkers, and controlled by an external oscillatory magnetic
field \cite{Dreyfus}. Another possibility is to take advantage of
{\em phoretic} effects, where gradients of fields such as
concentration, temperature, electric field, etc, couple to the
surface properties of particles to create slip velocity patterns
that could lead to net propulsion \cite{Anderson}.

A particularly appealing strategy for propelling small devices is to
take advantage of chemical reactions \cite{review}. In a pioneering
experiment, Paxton {\em et al.} observed autonomous motion of
platinum/gold nanorods \cite{paxton-1}, and similar experiments have
been performed by Fourneir-Bodoz et al who used gold/nickel nanorods
\cite{ozin} and Mano and Heller who used enzymes \cite{mano}. One
simple strategy for converting chemical energy to mechanical work in
such devices has been proposed by one of us and collaborators
\cite{prl-gla,janus-th}. In this scheme, a spherical particle is
considered with an asymmetric distribution of catalyst on its
surface.  If the chemical reaction so catalyzed produces more
products than it has reactants, then the asymmetric distribution of
reaction products propels the particle by a process of {\em
self-diffusiophoresis}. The experimentally realized swimmers
mentioned above are very similar in setup to the one proposed in
Ref. \cite{prl-gla}. However, it seems that the driving mechanism
for the propulsion in these experiment is different as it depends
strongly on the presence of a (bi-metal or bio-) electro-catalytic
structure \cite{paxton-2}. It is suggested that the pair of
electrochemical reactions at the two poles create a lateral electric
field near the particle surface (due to electron transfer) that
could move the solvent via electro-osmosis \cite{paxton-2}.

\begin{figure}[b]
\includegraphics[width=0.95\columnwidth]{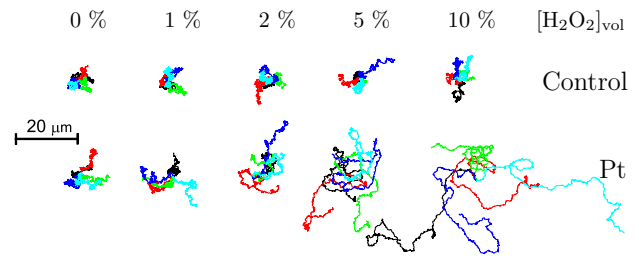}
\caption{(color online). Trajectories over 25 sec for $\times$5
particles of the control (blank) and Platinum-coated particles in
water, and varying solutions of hydrogen peroxide.}\label{fig:traj}
\end{figure}

Here we have realized the scheme proposed in Ref. \cite{prl-gla}
experimentally by taking polystyrene spheres with narrow size
distributions with a diameter of $1.62 \mu$m, and coating one side
of the spheres with platinum keeping the second half as the
non-conducting polystyrene. The platinum catalyzes the reduction of
a ``fuel'' of hydrogen peroxide to oxygen and water, which produces
more molecules of reaction product than consumed fuel. We fully
characterize the motion of the artificial micro-scale swimmer using
particle tracking, and probe the properties of the motion as a
function of hydrogen peroxide concentration. We show that at short
times, the particles move predominantly in a directed way, with a
velocity that depends on the concentration of the fuel molecules
with a Michaelis-Menten behavior. At longer times, the motion
reverts to a random walk, in which runs of directed motion are
interrupted by random changes of direction. We also extract the
positional and the rotational diffusion coefficients and show that
they are consistent with theoretical predictions, with the
rotational diffusion featuring a moderate concentration dependence
that could be attributed to directed rotational components in the
motion.

\begin{figure}
\includegraphics[width=0.8\columnwidth]{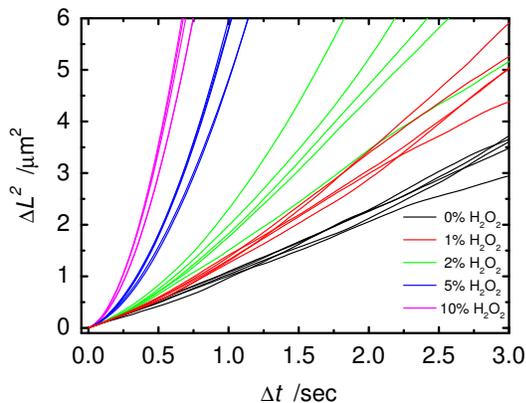}
\caption{(color). Mean squared displacement as a function of time
interval for the Pt-coated sphere trajectories shown in Fig.
\ref{fig:traj}, corresponding to different hydrogen peroxide
concentrations.}\label{fig:msd}
\end{figure}

Polystyrene microspheres were produced as described by Fujii et al
\cite{fuji}. The weight-average diameter of the microspheres was
determined by dynamic light scattering to be $1.62 \pm 0.13 \mu$m,
with a polydispersity index of 1.008. To half-coat the spheres
\cite{janus-exp} in platinum, a dilute suspension (0.05 wt\%) of the
spheres in isopropanol (HPLC grade: Fischer) was drawn over cleaned
glass microscope slides to produce a dilute monolayer of the spheres
($\sim 10^5$ spheres$\cdot {\rm cm}^{-2}$). A thin layer of Platinum
(99.99\% Agar Scientific) was evaporated onto the microsphere coated
slides resulting in a layer of Platinum 5.5 nm thick on one side of
the spheres. The half-platinum coated spheres were detached from the
glass slide using a sheet of PTFE (200 $\mu$m thick - Goodfellows)
acting as a blade and the detached spheres resuspended in 2 ml of
distilled water (18.2 M$\Omega$ cm$^{-1}$ - Purite). Hydrogen
peroxide solutions were prepared by subsequent dilutions of 30 \%
Hydrogen Peroxide solution (Perdrogen - Riedel-de Ha\"en). For a
given solution of hydrogen peroxide, 100 $\mu$l of the microsphere
containing stock solution was mixed with 2 ml of the appropriate
solution. A cuvette (1 mm optical path length - Hellma, cleaned with
Piranha Etch) was rinsed three times with the solution before
filling again and sealed using a PTFE stopper. Particle tracking was
achieved using a Nikon ME600 optical microscope mounted on an
isolation table and fitted with a Pixelink PL-A742 machine vision
camera, a $\times$20 objective (Nikon), and inverted illumination. A
movie with a field of view of approximately 120 $\mu$m $\times$ 120
$\mu$m of 3000 frames at a rate of 38 frames per second was recorded
for each particle. For each concentration of hydrogen peroxide
approximately 20 separate particles were tracked and analyzed using
a Labview (National Instruments) script providing particle
trajectory and timing (time, x/$\mu$m, and y/$\mu$m).

\begin{figure}
\includegraphics[width=0.75\columnwidth]{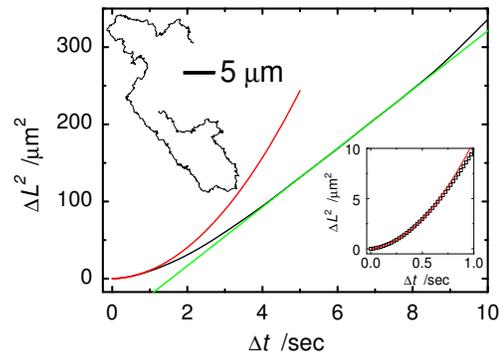}
\caption{(color online). Mean squared displacement as a function of
time for the inset trajectory (Pt-coated sphere in 10\% H$_2$O$_2$)
fitted to Eq. (\ref{deltaL2-t}), which resembles a parabola (red
line) at $\Delta t \ll \tau_{\rm R}$ and a straight line (green
line) at $\Delta t \gg \tau_{\rm R}$. The fits yield $v=3.1 \;\mu
{\rm m}\, {\rm s}^{-1}$, $D=0.31 \;\mu {\rm m}^2 {\rm s}^{-1}$, and
$\tau_{\rm R}=3.9$ s for this particular
trajectory.}\label{fig:fit-msd}
\end{figure}

\begin{figure*}
\includegraphics[width=1.99\columnwidth]{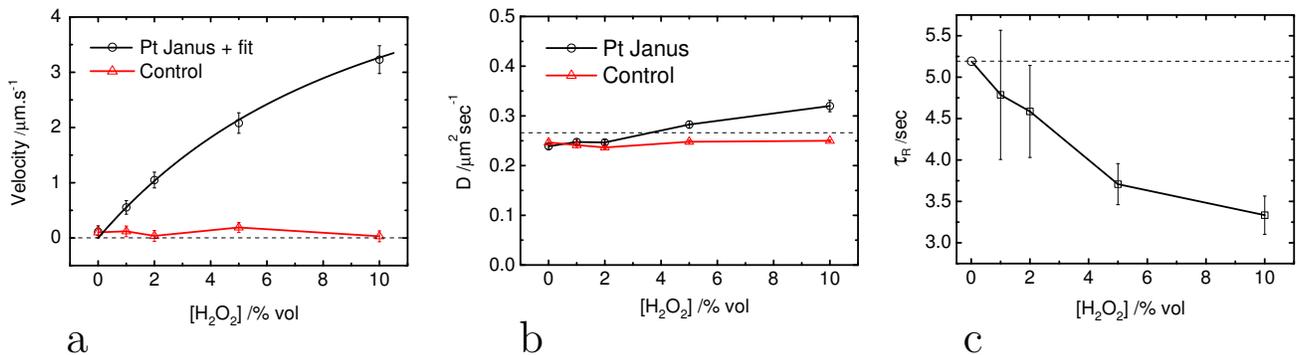}
\caption{(color online). (a) Velocity determined from the $\Delta t
\ll \tau_{\rm R}$ behavior for the control ($\vartriangle$) and
Pt-coated particles ($\circ$). The solid line is the line of best
fit using Eqs. (\ref{V-prop}) and (\ref{kMM}), with $a=1.2 \AA$,
$\lambda=5 \AA$, $k_1=4.4 \times 10^{11} \; \mu {\rm m}^{-2} {\rm
s}^{-1}$, and $k_2=4.8 \times 10^{10} \; \mu {\rm m}^{-2} {\rm
s}^{-1}$. (b) Diffusion coefficient determined from the $\Delta t
\ll \tau_{\rm R}$ behavior for the control and Pt-coated particles
with the theoretical value (-----) indicated. (c) Rotational
diffusion time determined at $\Delta t \gg \tau_{\rm R}$ (from the
gradient). Value at [H$_2$O$_2$]=0 is the theoretical
value.}\label{fig:VDTR}
\end{figure*}

Figure \ref{fig:traj} shows particle traces for a Pt-coated bead as
well as a control polystyrene bead. While the control particle
undergoes a characteristic Brownian motion, the Pt-coated particle
shows a systematic enhancement of a directional component in its
motion, as the concentration of hydrogen peroxide increases
\cite{epaps}. The motion can be analyzed more quantitatively using
particle tracking.  From the records of the particle trajectory, we
calculate the average value of the squared displacement as a
function of time. For a purely Brownian particle of radius $R$, the
squared displacement is linear in time with the slope controlled by
the particle diffusion coefficient $D=k_{\rm B} T/(6 \pi \eta R)$,
where $k_{\rm B}T$ is the thermal energy, $\eta$ is the viscosity of
water. The particle will also undergo rotational diffusion with a
characteristic (inverse) time scale $\tau_{\rm R}^{-1}=k_{\rm B}
T/(8 \pi \eta R^3)$ (also called the rotational diffusion
coefficient), with the two stochastic modes decoupled from each
other. For a particle propelled with velocity $V$, the direction of
motion is itself subject to rotational diffusion that leads to a
coupling between rotation and translation. In this case, one can
show that the (2D projection) mean squared displacement is given as:
\begin{equation}
\Delta L^2=4 D \Delta t+\frac{V^2 \tau_{\rm R}^2}{2}\left[\frac{2
\Delta t}{\tau_{\rm R}}+e^{-2 \Delta t/\tau_{\rm
R}}-1\right].\label{deltaL2-t}
\end{equation}
This expression has limiting forms of $\Delta L^2=4 D \Delta t+V^2
\Delta t^2$ for $\Delta t \ll \tau_{\rm R}$ and $\Delta L^2=(4 D+V^2
\tau_{\rm R}) \Delta t$ for $\Delta t \gg \tau_{\rm R}$.  At short
times, the contribution to the displacement due to the propulsion is
linear in time, while the Brownian displacement is proportional to
the square root of time, with these two contributions adding in
quadrature.  At times long compared to the rotational diffusion
time, rotational diffusion leads to a randomization of the direction
of propulsion, and the particle undergoes a random walk whose step
length is the product of the propelled velocity V and the rotational
diffusion time, leading to a substantial enhancement of the
effective diffusion coefficient over the classical value, namely
$D_{\rm eff}=D+\frac{1}{4}V^2 \tau_{\rm R}$. A similar crossover
behavior has been observed in the motion of tracer beads in the flow
caused near surfaces covered by bacterial carpets, as well as beads
propelled by adsorbed bacteria \cite{biophys-berg}.

Figure \ref{fig:msd} shows the average mean squared displacement as
a function of elapsed time for half-coated particles immersed in
hydrogen peroxide at various concentrations.  Each curve in Fig.
\ref{fig:msd} is produced by averaging over 3000 frames. In the
absence of hydrogen peroxide, the curves are linear, indicating the
simple diffusive behavior expected of a particle undergoing Brownian
motion. With the addition of hydrogen peroxide they show an
increasingly important parabolic component; the particles are being
propelled by the local osmotic pressure gradient created by the
asymmetric chemical reaction. Fitting these plots to the limiting
forms of the interpolation formula of Eq. (\ref{deltaL2-t}) yields
values for the parameters ($V$, $D$, and $\tau_{\rm R}$) as
functions of H$_2$O$_2$ concentration (see Fig. \ref{fig:fit-msd}).
For each concentration, we perform averaging over 20 independent
movies for the values of the fitting parameters.

Figure \ref{fig:VDTR}a shows the propulsion velocity as a function
of the hydrogen peroxide concentration. As we expect, as the
concentration of ``fuel'' rises the velocity with which the Pt
half-coated particles are propelled rises from the value of zero
observed in the absence of fuel; for the control particles, without
catalyst, no propulsion is observed. The propulsion velocity can be
calculated using the lateral gradient of the excess solute
concentration $C$ in the vicinity of the particle surface. This
gradient creates a slip velocity $V_s=\mu
\partial_{\parallel}C$ with the diffusiophoretic mobility given as
$\mu=k_{\rm B}T \lambda^2/\eta$ where $\lambda$ represents the range
of the interaction zone between the solute and the particle
\cite{Der}. One can then solve the Stokes hydrodynamics around the
sphere subject to this local slip velocity pattern on its surface,
and find the propulsion velocity. For a particle which is
half-coated with a material that can produce excess particles with
diffusion coefficient $D_o$ at a rate per unit area $k$, the
propulsion velocity is \cite{janus-th}:
\begin{equation}
V=\frac{\mu k}{4 D_o}=\frac{3\pi}{2} k a \lambda^2,\label{V-prop}
\end{equation}
where the hydrodynamic radius of the solute $a$ is introduced by way
of the Stokes-Einstein relation.

Since the velocity is directly proportional to the effective surface
reaction-rate $k$ [Eq. (\ref{V-prop})], the linear initial increase
and the subsequent tendency towards saturation observed in Fig.
\ref{fig:VDTR}a suggests that the Pt-catalyzed break-up of
H$_2$O$_2$ probably occurs in two stages of formation of a
Pt(H$_2$O$_2$) complex at a rate per unit area $k_1 [{\rm H}_2 {\rm
O}_2]_{\rm vol}$, in which $[{\rm H}_2 {\rm O}_2]_{\rm vol}$ has the
units of volume percentage, followed by a decomposition into water
and oxygen at a rate per unit area $k_2$. This leads to a
Michaelis-Menten behavior similar to enzymes \cite{h2o2}, as opposed
to the kinetics of decomposition in homogenous solution, which is
second order and would produce an upward curvature in Fig.
\ref{fig:VDTR}a. Solving the diffusion-reaction equations with the
proposed reaction kinetics, we find
\begin{equation}
k=k_2 \;\frac{[{\rm H}_2 {\rm O}_2]_{\rm vol}}{[{\rm H}_2 {\rm
O}_2]_{\rm vol}+k_2/k_1}.\label{kMM}
\end{equation}
The solid line in Fig. \ref{fig:VDTR}a is obtained from equations
(\ref{V-prop}) and (\ref{kMM}), with $a=1.2 \AA$, $\lambda=5 \AA$,
$k_1=4.4 \times 10^{11} \; \mu {\rm m}^{-2} {\rm s}^{-1}$, and
$k_2=4.8 \times 10^{10} \; \mu {\rm m}^{-2} {\rm s}^{-1}$, which are
reasonable numbers. In contrast to the strong dependence of the
propulsion velocity on hydrogen peroxide, the fitted particle (bare)
diffusion coefficient $D$ shown in Fig. \ref{fig:VDTR}b has little
dependence on the presence or absence of catalyst and the
concentration of hydrogen peroxide, and is close to the predicted
value from Einstein-Stokes relation (a slight residual dependence on
hydrogen peroxide concentration may be a result of a coupling
between these two fitting parameters in the analysis of the
displacement curves). Figure \ref{fig:VDTR}c shows fitted rotational
diffusion time. As mentioned above, the long time effective
diffusion is substantially enhanced due to the propulsion. For the
largest hydrogen peroxide concentration, we find a value of $D_{\rm
eff}=9.0 \;\mu {\rm m}^2 {\rm s}^{-1}$, which is an enhancement over
the purely Brownian value $D$ by a factor of nearly 30.

The rotational diffusion time shows a systematic decrease as a
function of hydrogen peroxide concentration for the coated
particles, with no dependence for the controls. This could be due to
the surface reaction imparting a small net angular velocity
$\omega$, and we surmise that it might be caused by inadvertent
asymmetric Pt coverage during the fabrication. Similar to the
translational diffusion coefficient, we would expect a combination
$\tau_{\rm R}^{-1}+\omega^2 \tau_{\rm R}$ to serve as the
renormalized rotational diffusion coefficient. This hypothesis is
strengthened by the observation of cycloid trajectories for some
samples at high H$_2$O$_2$ concentrations. At first sight this would
seem an unwelcome side effect---an increase in rotational velocity
has the effect of decreasing the renormalized diffusion coefficient
for the long-time behavior of the propelled particles.  But it also
raises the intriguing possibility of designing a system in which the
linear propulsive and rotational behavior are independently
controlled; this would constitute a system in which the step size of
a random walk could be controlled by an external parameter, opening
up the possibility of designing a system capable of chemotaxis, in a
way analogous to the bacteria {\em E. Coli} \cite{berg}.

By contrast to the existing experimental examples of autonomous
swimmers \cite{paxton-1,ozin,mano,paxton-2}, the propulsion
mechanism for our platinum/polystyrene particles does not involve
electrochemical reactions, and thus they realize a new class of
micro- and nano-scale chemical locomotion. We note that for this
class of (self-diffusiophoretic) swimmers, spherical geometry is
better than a rod geometry, because the velocity for rods is reduced
by a factor of the aspect ratio \cite{janus-th}. Moreover, their
true directional motion is easier to resolve experimentally as the
lower friction coefficient of the rods along their length causes
them to spend more time going along their own direction, which in
short time probing might be mistaken with directional locomotion.

In conclusion, we show that spatially asymmetric catalysis at the
surface of synthetic micron-scale particles can lead to effective
autonomous propulsion of spherical colloids. At short times, we
observe directed propulsion of the particles with velocities in the
$\mu {\rm m}/{\rm s}$ range; at longer times the direction of motion
of the particles is randomized, and the motion of the particles
becomes diffusive in character with an effective diffusion
coefficient which is substantially enhanced over the Brownian value.
We hope that our quantitative characterization of the motion of the
swimmer sheds some light on the fundamental issues involved in
designing chemical locomotive systems, and could inspire new
directions for their implementation.

\begin{acknowledgments}
R.G. acknowledges V. Sanei and M.T. Sarbolouki for their help during
the early stages of the experiment. This work was supported by the
EPSRC.
\end{acknowledgments}

%\begin{references}

\end{document}